\documentclass{svmult}

\usepackage{makeidx}  
\usepackage{graphicx}  
\usepackage{subfigure}   
\usepackage{mathptmx,helvet,courier}  
\usepackage{multicol}        
\usepackage[bottom]{footmisc} 
\usepackage{color}
\usepackage{amsmath}

\def\rcgindex#1{\index{#1}}
\def\myidxeffect#1{{\bf\large #1}}

\makeindex             

\begin{document}


 \title*{Extreme waves and branching flows in optical media} 
  \titlerunning{Extreme waves and branching flows in optical media}
 \author{M.~Mattheakis \and  
  G.P.~Tsironis
 } 
\institute{M.~Mattheakis \at Crete Center of Quantum Complexity and Nanotechnology (CCQCN), Department of Physics, University of Crete, Heraklion, Greece; and Institute of Electronic Structure and Laser, Foundation for Research and Technology - Hellas (FORTH), Heraklion, Greece\\ \email{mariosmat@physics.uoc.gr}
 \and G.P.~Tsironis
 \at
Crete Center of Quantum Complexity and Nanotechnology (CCQCN), Department of Physics, University of Crete, Heraklion, Greece; and Institute of Electronic Structure and Laser, Foundation for Research and Technology - Hellas (FORTH), Heraklion, Greece; and Department of Physics, Nazarbayev University, 53 Kabanbay Batyr Ave., Astana, 010000, Republic of Kazakhstan}
\authorrunning{M.~Mattheakis}

\maketitle


\abstract{
We address  light propagation properties in  complex media consisting of random
distributions of lenses that have specific focusing properties. We present both analytical and numerical techniques that
can be used to study emergent properties of light organization in these media. 
As light propagates,  it  experiences multiple
scattering leading to the formation of light bundles in the form of branches; these 
 are random yet occur systematically in the the medium, particularly
in the weak scattering limit. On the other hand, in the strong scattering limit
we find that coalescence of branches may lead to the formation of extreme waves
of the ``rogue wave'' type.  These waves appear at specific locations and 
arise in the linear as well as in the nonlinear regimes.  We present both the weak and strong scattering
limit and show that these complex phenomena can be studied numerically
and analytically through simple models.
 }

\section{Introduction}
\label{marios:sec:intro}


The propagation of waves in complex media is a currently topic of scientific interest with both theoretical and practical implications.  Wave phenomena abound in nature; for example, waves  at sea exhibit a plethora of wave phenomena, scaling up ranging from small amplitude ripples to larger (but still periodic) waves to gigantic and highly destructive solitary-like waves such as tsunamis and rogue waves.  The propagation properties of waves  is an important scientific problem  addressed b different means, namely, theoretical, numerical and experimental means.

  A complex medium gives rise to novel phenomena in wave propagation.  It is well known that  waves may interfere leading to local amplitude enhancement or diminution.  This feature may be amplified by the properties of the complex medium leading to very large transients as well as non-uniform propagation.  The resulting complex dynamics is reminiscent to
 similar phenomena that appear in condensed mater physics and other areas.  In the present Chapter, we will focus primarily on two dimensional wave evolution in several types of random media.  The unifying feature is that the wave scatterers have specific properties affecting drastically the wave propagation.  They typically focus or defocus strongly the waves leading to  phenomena with spatiotemporal complexity.  Specifically, as the wave propagates there are bifurcations in space leading to
a light flow that is split in dominant as well as in smaller branches.  When the propagation of an electromagnetic wave takes place in a dielectric, we observe dominant channels of wave coalescence ``decorated'' with smaller and smaller channels, which constitute a typical fractal-like picture in space.  Light propagation becomes  complex as a result of the strong but random influence of the medium on the dynamics.

In addition to branching we may have other effects derived from the enhanced but random focusing and interference.  In specific locations of the medium a giant fluctuation may appear generating a spatiotemporal  ``hot spot''.  These transients may classify as optical rogue
or freak waves, similar in several ways to the ones that appear at sea.  The latter are giant waves appearing essentially ``from nowhere''
 (while the oceanic conditions are not necessarily very bad).  There are numerous reports on  rogue waves that carry  high energy and are  destructive for ships and lives. In the optics context, these extreme waves seem to depend very much both on the randomness of the medium and on its strong focusing or defocusing properties (strong scattering random medium).  Thus, the complexity features of electromagnetic wave
propagation in the medium consists of both branching aspects and rogue wave formation.  They seem to originate from a similar source, although  light branches appear also when the medium is weakly disordered.

The structure of this Chapter is as follows. In section \ref{marios:sec:mathTools} we present  the mathematical and computational methods that have been used for the investigation of the electromagnetic wave propagation; in particular, emphasis is placed on the geometrical optics limit and on the Finite Difference in Time Domain (FDTD)  method. The latter is a basic technique used for the numerical solution of Maxwell Equations. These methods are used in order to investigate the electromagnetic wave propagation through certain configurations of
special lenses known as Luneburg lenses. In Section \ref{marios:sec:branching} we explore the electromagnetic wave propagation through a weak scattering random medium and present findings on the appearance of caustic formation of light rays. Caustics are studied by means of the Lagrangian manifold method and a statistical scaling law; the latter determines the position where the first caustic appears. In Section \ref{marios:sec:RWs} we investigate electromagnetic wave propagation through a strong scattering random medium and show that rogue waves can emerge in such systems even when nonlinearity is absent. Finally, in Section \ref{marios:sec:conclusion} we conclude and  present a summary of the findings.

\section{Mathematical tools for electormagnetic wave propagation propagation}
\label{marios:sec:mathTools}

 In this Section, we present methods that can be used in order to determine the characteristics of  light propagation in an inhomogeneous isotropic medium.  We consider structures embedded in the medium that have cylindrical symmetry and are described via the  refractive index $n(r)=\sqrt{\epsilon}$, where $r$ is the radial coordinate of the structure. We focus on a propagating electromagnetic field   near the visible spectrum; in this regime, light oscillates very rapidly (with frequencies of the order of $10^{14}$ Hz) resulting in very large magnitudes of the wavevector (i.e. $k\rightarrow\infty$) and very small magnitudes of wavelength ($\lambda\rightarrow 0$). In this limit, the wave behaviour of light can be neglected and the optical laws can be formulated in geometrical terms, i.e., the electromagnetic waves are treated as rays. This approximation is well known  as geometrical optics \rcgindex{\myidxeffect{G}!Geometrical optics}
 and holds as the size of lenses (or obstacles) in the medium are much larger than the wavelength \cite{bornBook, klineBook,lagrangianOptics,luneburgBook}.

To follow the electromagnetic wave evolution in this inhomogeneous medium, one may
use one of three methods of geometrical optical propagation, which will be  outlined in detail below.  These methods will be applied specifically in a medium comprising spatial distributions of Luneburg lenses  \rcgindex{\myidxeffect{L}!Luneburg lens} \cite{luneburgBook,marios2012}. The Luneburg lens is a spherical lens with index of refraction that varies radially from the value one $(r=1)$ in the outer boundary (when the surrounding medium is vacuum or air)  to $r=\sqrt{2}$ in the center ; the functional dependence of the index of refraction on the radius is given by:
 \begin{equation}
n(r)=\sqrt{2-\left(\frac{r}{R}\right)^2}
 \label{marios:eq:luneburgIndex} 
 \end{equation}
 where $R$ is the radius of the Luneburg lens. The basic property of a Luneburg lens  is that,  in the geometrical optics limit,  parallel rays impinging on the spherical surface are focused to the opposite side of the lens. This feature makes Luneburg lenses quite interesting for applications since the focal surface is predefined for parallel rays of any initial angle. Luneburg lenses can be used to form   gradient index (GRIN)  optical metamaterials \rcgindex{\myidxeffect{M}!metamaterials}; in the latter one exploits the spatial variation of the index of refraction in order to enhance light manipulation in a variety of circumstances \cite{marios2012}.  In the specific analysis, which will follow, we use primarily two dimensional media and, as a result, we will employ cylindrical Luneburg lenses that have however the same index variability as the one of 
 equation (\ref{marios:eq:luneburgIndex}), with $r$ the radial coordinate and $R$ the radius of the cylindrical lens.

We  apply three distinct geometrical optics methods in order to analyze light propagation.  The first is based 
on  Fermat's principle that optimizes the optical path traversed by light and  by means of it, an exact ray tracing equation for a single Luneburg lens is derived (subsection \ref{marios:subsec:quasi2D}); this approach is essentially a quasi two dimensional approximation.  The second method is a parametric  two dimensional method based also on Fermat's principle; in this method,   the  arc length $s$  in the light trajectory is used as a free parameter (subsection \ref{marios:subsec:rayParametric}).  The third  geometrical optics approach is based on  the Helmholtz wave equation (subsection \ref{marios:subsec:helmholtzRays}). The results of the these three geometrical optics methods are compared in subsection \ref{marios:subsec:fdtd} along with the corresponding numerical solution of the time dependent Maxwell equations through the Finite Difference in Time Domain  method.

\subsection{Quasi-two dimensional ray solution}
\label{marios:subsec:quasi2D}

The time $T$ that light takes to traverse a path between two points $A$ and $B$ in space is given by the integral \cite{bornBook, stauroudisBook}
\begin{equation}
\label{marios:eq:fermatTime}
T=\int_A^B dt =\frac{1}{c}\int_A^B n ds 
\end{equation}
where the infinitesimal time $dt$ can be written in arc length terms as $dt=ds/v$ and $v$ is the velocity of light in a medium with refractive index $n$ $\left(v=c/n\right)$, where $c$ the velocity of light in the bulk medium. 

In the special case where the medium has spherical or cylindrical symmetry and thus $n(\vec{r})\equiv n(r)$, the optical path length $S$ of a ray propagating from point $A$ to  point $B$ is 
 \cite{bornBook,klineBook,lagrangianOptics,luneburgBook,marios2012,stauroudisBook}
\begin{equation}
 \label{marios:eq:fermatPrinciple} 
S=\int_A^B n(r) ds
\end{equation}
In polar coordinates, the arc length is $ds=\sqrt{dr^2+r^2 d\phi^2}$, where $r$, $\phi$ are the radial and angular polar coordinates respectively. In the quasi two dimensional  approximation  \rcgindex{\myidxeffect{Q}!Quasi 2D approximation} the coordinate $r$ can be  considered as ``generalized'' time and therefore the arc length can be written as $ds=\sqrt{1+r^2\dot\phi^2}dr$, with  $\dot \phi \equiv d\phi / dr$. As a result, the Fermat's variational integral \rcgindex{\myidxeffect{F}!Fermat principle} of equation (\ref{marios:eq:fermatPrinciple}) becomes
 \begin{equation}
 \label{marios:eq:fermat2} 
S=\int_A^B n(r)\sqrt{1+r^2\dot{\phi}^2} dr
\end{equation}
yielding the optical Lagrangian \rcgindex{\myidxeffect{O}!Optical Lagrangian}
 \cite{lagrangianOptics,luneburgBook,marios2012,stauroudisBook}
\begin{equation}
 \label{marios:eq:lagrangian1} 
L(\phi,\dot \phi, r)=n(r)\sqrt{1+r^2 \dot \phi^2} 
\end{equation}
The shortest optical path is obtained via the minimization of the integral in equation (\ref{marios:eq:fermat2}) and can be calculated by solving  the Euler-Lagrange equation for the Lagrangian of equation (\ref{marios:eq:lagrangian1}), viz.
\begin{equation}
\frac{d}{dr} \frac{\partial L}{\partial \dot{\phi}}=\frac{\partial L}{\partial \phi}
\end{equation}
Since the Lagrangian \rcgindex{\myidxeffect{O}!Optical Lagrangian} of equation (\ref{marios:eq:lagrangian1}) is cyclic in $\phi$,
$\partial {L} / \partial \phi=0$ and, thus, $\partial {L} / \partial \dot{\phi}=C$
where $C$ is a constant. The resulting equation of motion \cite{lagrangianOptics,marios2012,stauroudisBook} 
\begin{equation}
\label{marios:eq:diff1st}
\frac{n(r)r^2 }{\sqrt{1+r^2 \dot{\phi}^2}}\dot{\phi}=C
\end{equation}
is a nonlinear differential equation describing the trajectory $r(\phi )$ of a ray in an isotropic medium with radial symmetry and refractive index $n(r)$. Replacing the term $\dot \phi\equiv d\phi / dr$ and solving for $d\phi$, we obtain a first integral of motion \cite{bornBook,lagrangianOptics,marios2012}, that is
\begin{equation}
\label{marios:eq:firstIntegral}
\int{d\phi} = \int{\frac{C}{r \sqrt{n^2r^2-C^2}}dr}
\end{equation}
Equation (\ref{marios:eq:firstIntegral}) holds for arbitrary refractive indexes $n(r)$.  The differential equation (\ref{marios:eq:diff1st}) and the integral (\ref{marios:eq:firstIntegral}) are the most important results of this subsection; they provide, for a specific refractive index profile, namely, the ray tracing equation \rcgindex{\myidxeffect{R}!Ray tracing}
 for $r(\phi)$.

In the specific case of a single Luneburg lens with  the refractive index function of equation (\ref{marios:eq:luneburgIndex}), the
ray tracing solution in its interior is written as:
\begin{equation}
\label{marios:eq:rayeq1}
r(\phi) =  \frac{{C'} R}{ \sqrt{1-\sqrt{1-C'^2}~\sin\left(2(\phi+\beta)\right)} }
\end{equation}
where $C'$ and $\beta$ are constants. This analytical expression may be cast in a direct Cartesian
form for the $(x,y)$ coordinates of the ray; after some algebra we obtain
\begin{eqnarray}
\label{marios:eq:ellipsis1}
\left(1-T\sin(2\beta)\right)x^2+\left(1+T\sin(2\beta)\right)y^2 
-2T\cos(2\beta) xy+\left(T^2-1\right)R^2=0
\end{eqnarray}
where $T$ and $\beta$ are constants. We note that equation (\ref{marios:eq:ellipsis1}) is the equation of an ellipse. This  result agrees  with the Luneburg theory and states that inside a Luneburg lens light follows elliptic orbits \cite{luneburgBook,marios2012}.

\begin{figure}[h] \begin{center}
\includegraphics[scale=.30]{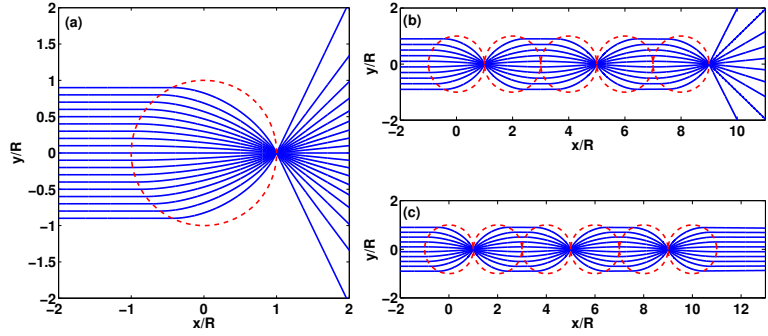}
 \caption{ The  dashed lines denote the arrangement of the Luneburg lenses. The solid lines represent light rays that have been computed by the analytical ray tracing equation (\ref{marios:eq:generalSolution1}). (a) Ray tracing through a single Luneburg lens; all rays are focused on a single point. (b, c) Light is guided by Luneburg lenses across the linear network constituted of  five Luneburg lenses in (b) and of six Luneburg lenses in (c).}
\label{fig:fig1}
\end{center}
\end{figure}

The constants $T$ and $\beta$  of equation (\ref{marios:eq:ellipsis1}) are determined by the ray boundary (or the ``initial" conditions and depend on the initial propagation angle $\theta$ of a ray that enters the lens at the point  $(x_0,y_0)$ located on the circle at the lens radius $R$ \cite{lagrangianOptics,marios2012}. The entry point of the ray is at $(x,y)=-R(\cos\theta , \sin\theta )$. Substituting these expressions in equation (\ref{marios:eq:ellipsis1}) we obtain 
\begin{equation}
\label{marios:eq:Tconst1}
T=\sin\left(2\beta +2\theta\right)
\end{equation}
In order to determine the constants $T$ and $\beta$, we need an additional relation connecting them. We take the derivative of the equation (\ref{marios:eq:ellipsis1}) with respect to  $x$ and utilize the relation $dy/dx=\tan(\theta)$, where $\theta$ the initial propagation angle. In addition, using  $(x_0,y_0)$ for the initial ray point on the Luneburg lens surface, we set $x=x_0$ and $y=y_0$ in  equation (\ref{marios:eq:ellipsis1}) and solve for $T$, getting
 \begin{equation}
\label{marios:eq:Tconst2}
 T = \frac{x_0+y_0 \tan(\theta)}{\tan(\theta)\left[x_0 \cos(2\beta)-y_0\sin(2\beta)\right]+\left[x_0 \sin(2\beta)+ y_0 \cos(2\beta)\right]} 
\end{equation} 
Equations (\ref{marios:eq:Tconst1}) and (\ref{marios:eq:Tconst2}) comprise an algebraic nonlinear system expressing the constants $T$ and $\beta$ as a function of the initial ray entry point in the Luneburg lens at $(x_0 , y_0 )$ with initial propagation angle $\theta$.  Combining  equations (\ref{marios:eq:Tconst1}) and (\ref{marios:eq:Tconst2}) we obtain 
\begin{equation} 
\label{marios:eq:betaConst}
\beta=\frac{1}{2}\left(\tan^{-1}({x_0}/{y_0})-\theta\right)
\end{equation}
therefore, according to equation (\ref{marios:eq:Tconst1})
\begin{equation}
\label{marios:eq:TConst}
T=\sin\left(\tan^{-1}(x_0 /y_0 )+\theta\right)
\end{equation} 
Substituting now equations (\ref{marios:eq:betaConst}) and (\ref{marios:eq:TConst}) to equation (\ref{marios:eq:ellipsis1}) and solving for $y$, we obtain the ray tracing equation \rcgindex{\myidxeffect{R}!Ray tracing}
\begin{eqnarray}
\label{marios:eq:generalSolution1}
y(x) & =  &  \frac{ \left(2x_0 y_0+ R^2 \sin(2\theta)\right)}{2x_0^2+\left(1+\cos(2\theta)\right)R^2}~x\nonumber \\ 
& & {} + \frac{\sqrt{2}R y_0\cos(\theta)\sqrt{\left(1+\cos(2\theta)\right)R^2+2x_0^2-2x^2} }{2x_0^2+\left(1+\cos(2\theta)\right)R^2} \\
& & {} - \frac{x_0 \sin(\theta)\sqrt{\left(1+\cos(2\theta)\right)R^2+2x_0^2-2x^2}}{2x_0^2+\left(1+\cos(2\theta)\right)R^2} \nonumber
\end{eqnarray}
Equation (\ref{marios:eq:generalSolution1}) describes the complete solution of the ray trajectory through an Luneburg lens. 
In the simple case where all the rays are parallel to the $x$ axis and the initial angle  $\theta=0$,  equation (\ref{marios:eq:generalSolution1}) simplifies to
\begin{equation}
\label{marios:eq:LLsolution1}
y(x)=\frac{y_0}{x_0^2+R^2}\left(x_0 x+R\sqrt{R^2+x_0^2-x^2}\right)
\end{equation}
We note that in order to determine the exit angle $\theta '$, i.e. the angle with  which each ray exits the lens,
we take the arc tangent of the derivative of equation (\ref{marios:eq:generalSolution1}) with respect to $x$, at the  focal point on the surface of lens at $x=R\cos(\theta)$.  The solution of equation (\ref{marios:eq:generalSolution1}) can be used to study several configurations of Luneburg lenses. We present, in Fig.\ref{fig:fig1},  the ray tracing propagation based on the equation (\ref{marios:eq:generalSolution1}), through a single Luneburg lens \rcgindex{\myidxeffect{L}!Luneburg lens} (Fig.\ref{fig:fig1}a) and through two linear Luneburg lens waveguide \rcgindex{\myidxeffect{W}!Waveguide} networks (Fig.\ref{fig:fig1}(b,c)) \cite{marios2012}; in all cases the bulk media is air with refraction index $n_{air}=1$. 

When the rays  are scattered backwards, the quasi-two dimensional approximation \rcgindex{\myidxeffect{Q}!Quasi 2D approximation}  breaks down and the solution of equation (\ref{marios:eq:generalSolution1}) becomes complex. This failure is due to the assumption that the radial coordinate plays the role of time, viz. a monotonically increasing parameter similar to physical time.
In order to address light back-propagation it is advantageous  to use parametric solutions where the Cartesian ray coordinates $x,y$ are 
both time-dependent variables.  This approach is explained in subsections \ref{marios:subsec:rayParametric} and \ref{marios:subsec:helmholtzRays} where the parametric ray tracing solution is derived.

 
\subsection{Parametric two dimensional ray solution}
\label{marios:subsec:rayParametric}

Since the quasi two dimensional approximation \rcgindex{\myidxeffect{Q}!Quasi 2D approximation}
 fails for backscattered rays, we need to develop a real two dimensional parametric ray tracing equation. This is done through the use of  Fermat's principle \rcgindex{\myidxeffect{F}!Fermat principle} while assuming that both ray coordinates are time-dependent variables.

We use the infinitesimal arc length $ds=\sqrt{dx^2+dy^2}$ in Cartesian coordinates and further introduce the parameter $\tau$ as generalized time i.e.  $ds=\sqrt{\dot x^2+\dot y^2}~d\tau$ where the dot indicates differentiation with respect to parameter $\tau$, ($\dot \alpha \equiv da/d\tau$) and  $x\equiv x(\tau)$, $y\equiv y(\tau)$ \cite{klineBook,luneburgBook,marios2012,stauroudisBook}. The Fermat integral of equation (\ref{marios:eq:fermatPrinciple}) becomes
\begin{equation}
S=\int_A^B n(x,y)\sqrt{\dot x^2+\dot y^2}d\tau
\end{equation}
where $n(x,y)$ is the refractive index in Cartesian coordinates; Minimization of the travel path $S$ leads to the optical Lagrangian \rcgindex{\myidxeffect{O}!Optical Lagrangian}
\begin{equation}
\label{marios:eq:lagrangian2.2}
L(x,y,\dot x,\dot y,\tau)=n(x,y)\sqrt{\dot x^2+\dot y^2}
\end{equation}
We introduce the generalized optical momenta $k_x$, $k_y$ that are conjugate to $x$, $y$ represented as:
\begin{equation}
\label{marios:eq:kx}
k_x=\frac{\partial L}{\partial \dot x}=\frac{n \dot x}{\sqrt{\dot x^2+\dot y^2}}
\end{equation}
\begin{equation}
\label{marios:eq:ky}
k_y=\frac{\partial L}{\partial \dot y}=\frac{n \dot y}{\sqrt{\dot x^2+\dot y^2}}
\end{equation}
Equations (\ref{marios:eq:kx}) and  (\ref{marios:eq:ky}) comprise  an algebraic nonlinear system, which leads to
\begin{equation}
\label{marios:eq:momenta1}
k_x^2+k_y^2-n(x,y)^2=0
\end{equation}
We can rewrite  equation (\ref{marios:eq:momenta1})  in vector form using $\vec{r} \equiv (x,y)$ and $\vec{ k}\equiv (k_x,k_y)$, as
\begin{equation}
\label{marios:eq:momenta2}
\vec k^2-n(\vec{r})^2=0
\end{equation}
Multiplying  equation (\ref{marios:eq:momenta2}) with the factor $1/2$ reveals the direct  analogy to the equations of classical mechanics. The first term is the kinetic energy of the rays 
\begin{equation}
\label{marios:eq:kineticEnergy}
T=\frac{\vec k^2}{2}
\end{equation}
the second  is the corresponding potential energy  given by
\begin{equation}
\label{marios:eq:potentialEnergy}
V=-\frac{n(\vec r)^2}{2}
\end{equation}
while the total energy is given by
\begin{equation}
\label{marios:eq:Hamiltonian2}
H(\vec{r},\vec{k})=\frac{\vec k^2}{2}-\frac{n(\vec r)^2}{2}=0
\end{equation}
Physically, equations (\ref{marios:eq:kineticEnergy})-(\ref{marios:eq:Hamiltonian2}) represent the motion of a classical particle of unit mass under the influence of the potential $V( \vec r )$, while the total energy of the system is taken to be zero \cite{bornBook,klineBook,luneburgBook,marios2012}.

We can obtain a Hamiltonian ray tracing system by solving Hamilton's equations for the Hamiltonian \rcgindex{\myidxeffect{O}!Optical Hamiltonian}  of equation  (\ref{marios:eq:Hamiltonian2}) \cite{marios2012,oreficeBook,stauroudisBook}; we get
\begin{equation}
\label{marios:eq:rayEquation2a}
\frac{d\vec r}{d\tau}=\frac{\partial H}{\partial \vec k}=\vec k
\end{equation}
and
\begin{equation}
\label{marios:eq:rayEquation2b}
\frac{d\vec k}{d\tau}=-\frac{\partial H}{\partial \vec r}=\frac{1}{2}\nabla n(\vec r)^2
\end{equation}
where $\nabla \equiv \left(\frac{\partial}{\partial x},\frac{\partial}{\partial y} \right)$, $\tau$ is an effective time   related to real travel time $t$  through $d\tau=c~dt$, whereas $c$ is the velocity of rays in the bulk medium with index of refraction $n_0$ $\left( c=c_0/n_0 \right)$. Combining  equations (\ref{marios:eq:rayEquation2a}) and (\ref{marios:eq:rayEquation2b}) we obtain \cite{bornBook,klineBook,marios2012,stauroudisBook,luneburgBook,oreficeBook}
\begin{equation}
\label{marios:eq:eqmotion2a}
\frac{d^2{\vec r}}{d\tau^2}=\frac{1}{2} \nabla n(\vec r)^2 
\end{equation} 
and  restoring the real travel time $t$ instead of the effective time $\tau$, we obtain
\begin{equation}
\label{marios:eq:eqmotion2b}
\ddot{\vec r}=\frac{c^2}{2} \nabla n(\vec r)^2 
\end{equation}
where derivatives are taken with respect to travel time $t$, namely $\dot q=dq/dt$ for arbitrary $q(t)$. We conclude that  equation (\ref{marios:eq:eqmotion2b}) is a general equation of motion for ray paths in a medium with an arbitrary refractive index function $n(\vec r)$. The explicit solution for Luneburg lens will be given in the following  subsection.


\subsection{Helmholtz wave equation approach}
\label{marios:subsec:helmholtzRays}

An alternative geometrical optics approach \rcgindex{\myidxeffect{G}!Geometrical optics} may be developed starting from  the Helmholtz wave equation\rcgindex{\myidxeffect{H}!Helmholtz  equation}. In this approach we recover once again the ray tracing equation \rcgindex{\myidxeffect{R}!Ray tracing} (\ref{marios:eq:rayEquation2b}) and  find an explicit ray solution for light propagation through a Luneburg lens with refractive index  given by equation (\ref{marios:eq:luneburgIndex}).

A  monochromatic electromagnetic wave propagating in a two dimensional medium can be described by the  Helmholtz equation   \cite{klineBook,stauroudisBook}.
\begin{equation}
\label{marios:eq:helmholtz}
\left[\vec \nabla^2+\left(n k_0 \right)^2 \right]u(x,y)=0
\end{equation}
where $\nabla^2 = \frac{\partial^2}{\partial x^2}+\frac{\partial^2}{\partial y^2}$ is the Laplacian in a two dimensional space, and $u(x,y)$ is a scalar function representing any component of the electric or magnetic field; $n$ is the refractive index that generally depends on position, $k_0\equiv \omega/c=2\pi/\lambda_0$ is the wave vector in the bulk media whereas $\omega$ and $\lambda_0$ are the angular frequency and wavelength of the electromagnetic wave respectively and $c$ the velocity of the light \cite{klineBook,marios2012,stauroudisBook,oreficeBook}.
Although equation (\ref{marios:eq:helmholtz}) is time-independent and therefore we cannot investigate dynamical phenomena, we can determine the stationary paths followed by the light rays; this is known as the ray tracing approximation\rcgindex{\myidxeffect{R}!Ray tracing}. 

Assuming that the scalar field $u$ can be determined by an amplitude real function $A(x,y)$ and a phase  $\phi(x,y)$  real function (Sommerfeld-Runge assumption), where $\phi (x, y) $ is  known as the eikonal function  \rcgindex{\myidxeffect{E}!Eikonal function}
 we proceed with the well known transformation
\begin{equation}
\label{marios:eq:sommerfeld}
u(x,y)=A(x,y) e^{i\phi(x,y)}
\end{equation}
Substituting equation (\ref{marios:eq:sommerfeld})  into  the wave equation (\ref{marios:eq:helmholtz}) and separating the real from the imaginary parts, we obtain the following system of equations \cite{marios2012,oreficeBook}
\begin{gather}
\label{marios:eq:sys3a}
\left(\vec \nabla \phi \right)^2-\left( n k_0 \right)^2 =\frac{\nabla^2 A}{A} \\
\label{marios:eq:sys3b}
\vec \nabla \cdot \left(A^2\vec \nabla \phi\right)=0 
\end{gather}
Equation (\ref{marios:eq:sys3b}) expresses the constancy of the flux of the vector $A^2\vec \nabla \phi$ along any tube formed by the field lines of the wavevector defined through $\vec k=\vec \nabla  \phi$; the latter transforms  equation (\ref{marios:eq:sys3a}) into
\begin{equation}
\label{marios:eq:eikonalEq2}
\vec{k}^2-\left(n k_0\right)^2 =\frac{\nabla^2 A}{A}
\end{equation}
The last term in the equation (\ref{marios:eq:sys3a}), viz. $\frac{\nabla^2 A}{A}$ is the Helmholtz potential\rcgindex{\myidxeffect{H}!Helmholtz potential} \cite{marios2012,oreficeBook}; it preserves the wave behaviour in the ray tracing equation. In the geometrical optics \rcgindex{\myidxeffect{G}!Geometrical optics} limit where the space variation $L$ of the beam amplitude $A$ satisfies the condition $k_0 L >> 1$, i.e.  $\lambda<<L$, the Helmholtz potential vanishes; in this case equation (\ref{marios:eq:eikonalEq2}) gives the well known eikonal equation\rcgindex{\myidxeffect{E}!Eikonal function}, (equation (\ref{marios:eq:eikonalEq3})) \rcgindex{\myidxeffect{E}!Eikonal equation} which is the basic equation in the geometrical optics approach \cite{bornBook,klineBook,luneburgBook,oreficeBook,stauroudisBook}, viz.
\begin{equation}
\label{marios:eq:eikonalEq3}
\left(\vec \nabla \phi \right)^2=\left( n k_0 \right)^2 
\end{equation}
The most important result of this approach is that rays are not coupled any more and they propagate independently from one another.

 We can introduce the optical Hamiltonian by multiplying equation (\ref{marios:eq:eikonalEq2}) with the factor $c/(2k_0)$; this leads to
\begin{equation}
\label{marios:eq:hamiltonian3}
H(\vec r, \vec k)=\frac{c}{2k_0}\vec k^2-\frac{c k_0}{2}n^2(\vec r)
\end{equation}
Finally, the system of equations of motion can be written as a second order ordinary differential equation  by solving Hamilton's equation described by equations (\ref{marios:eq:rayEquation2a})(\ref{marios:eq:rayEquation2b}) and yields the same equation of motion found in the expression of equation (\ref{marios:eq:eqmotion2b}), viz. the equation
\begin{equation}
\label{marios:eq:rayTracing3}
\ddot{\vec r}=\frac{c^2}{2}\nabla n^2
\end{equation}
Substituting the Luneburg lens \rcgindex{\myidxeffect{L}!Luneburg lens} refractive index equation (\ref{marios:eq:luneburgIndex}) in the differential equation (\ref{marios:eq:eqmotion2b}), or equation (\ref{marios:eq:rayTracing3}), we obtain the following equation of motion   describing the ray paths inside a Luneburg lens:
\begin{equation}
\label{marios:eq:luneburgODE}
\ddot{\vec r}+\frac{c^2}{R^2}\vec r=0
\end{equation}
We may now proceed with the solution of the  equation (\ref{marios:eq:luneburgODE}). Using the boundary conditions $\vec r(0)=\vec r_0=(x_0,y_0)$ and $\dot{\vec r_0}=\vec k_0=(k_{0x},k_{0y})$ we obtain
\begin{equation}
\label{marios:eq:luneburgRay3}
\left( \begin{array}{c}  x(t) \\ y(t)  \end{array} \right) =
\left( \begin{array}{c}  x_0 \\  y_0  \end{array} \right) \cos\left(\frac{c}{R}t\right)+
\left( \begin{array}{c}  k_{0x} \\  k_{0y}  \end{array} \right) \frac{R}{c} \sin\left(\frac{c}{R}t\right)
\end{equation}
The solution (\ref{marios:eq:luneburgRay3}), in Cartesian coordinates, describes elliptical orbits, in agreement with Luneburg's theory \cite{luneburgBook} as well as with equations (\ref{marios:eq:generalSolution1}), (\ref{marios:eq:LLsolution1}) \cite{marios2012}.

\begin{figure}[h] \begin{center}
\includegraphics[scale=.40]{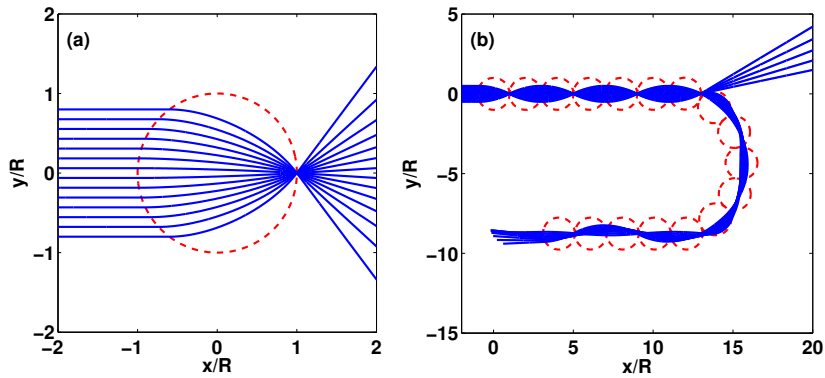}
 \caption{ The  dashed lines denote the arrangement of Luneburg lenses which are  spherical lenses
with index of refraction given by the equation (\ref{marios:eq:luneburgIndex}). The solid lines show the ray tracing performed via the analytical parametric solution of equation (\ref{marios:eq:luneburgRay3}). (a) Ray tracing through a single Luneburg lens, all rays are focused on a single point. (b) An $180^o$ reversed bend waveguide formed through seventeen Luneburg lenses. 
}
\label{fig:fig2}
\end{center}
\end{figure}

In Fig. \ref{fig:fig2} we present results based on the explicit ray solutions of equation (\ref{marios:eq:luneburgRay3}). The ray tracing\rcgindex{\myidxeffect{R}!Ray tracing} propagation through a single Luneburg lens is indicated  in Fig. \ref{fig:fig2}a; this is in agreement with the quasi two dimensional ray solution  shown in Fig. \ref{fig:fig1}a. In Fig. \ref{fig:fig2}b we show an $180^o$ reversed bend waveguide \rcgindex{\myidxeffect{W}!Waveguide} formed by seventeen Luneburg lenses; as can be seen, backward propagation can be described via the parametric ray solution.

\subsection{Numerical solution of Maxwell equations}
\label{marios:subsec:fdtd}

The Finite Difference in Time Domain (FDTD)\rcgindex{\myidxeffect{F}!Finite Difference in Time Domain} method  \rcgindex{\myidxeffect{F}!Finite Difference in Time Domain} is a  numerical method used in computational electrodynamics;  while most numerical methods are applied in the frequency domain, the Finite Difference in Time Domain method solves the time dependent Maxwell equations in the time domain, viz. the calculation of the electromagnetic field as it progresses at discrete steps both in time and space. Since it is a time domain method, the solutions of  Finite Difference in Time Domain can cover a wide frequency range with a single simulation.
The Finite Difference Time Domain \rcgindex{\myidxeffect{F}!Finite Difference in Time Domain} method is used in several  scientific and engineering problems related
to electromagnetic wave propagation and detection, such as  antennas, radiation and microwave applications, as well as in the interaction of electromagnetic waves with solid state structures such as in plasmonics and photonic crystals.

We apply the Finite Difference in Time Domain method for a monochromatic electromagnetic  plane wave source with wavelength $\lambda$, with  vacuum as the bulk material and  permittivity $\epsilon=1$. We use Luneburg lens with radius $ R=10\lambda$ and  permittivity based on equation (\ref{marios:eq:luneburgIndex}) i.e. $\epsilon = n^2=2-(r/R)^2$. We simulate electromagnetic propagation through a single Luneburg lens as in Fig. \ref{fig:fig1}a and Fig. \ref{fig:fig2}a, through a linearly spaced Luneburg lens waveguide system \rcgindex{\myidxeffect{W}!Waveguide} comprising of five Luneburg lenses, as in Fig. \ref{fig:fig1}b, and also through a $180^o$ reversed bend through a  waveguide formed by seventeen Luneburg lenses, as in Fig. \ref{fig:fig2}b.

\begin{figure}[h] \begin{center}
\includegraphics[scale=.35]{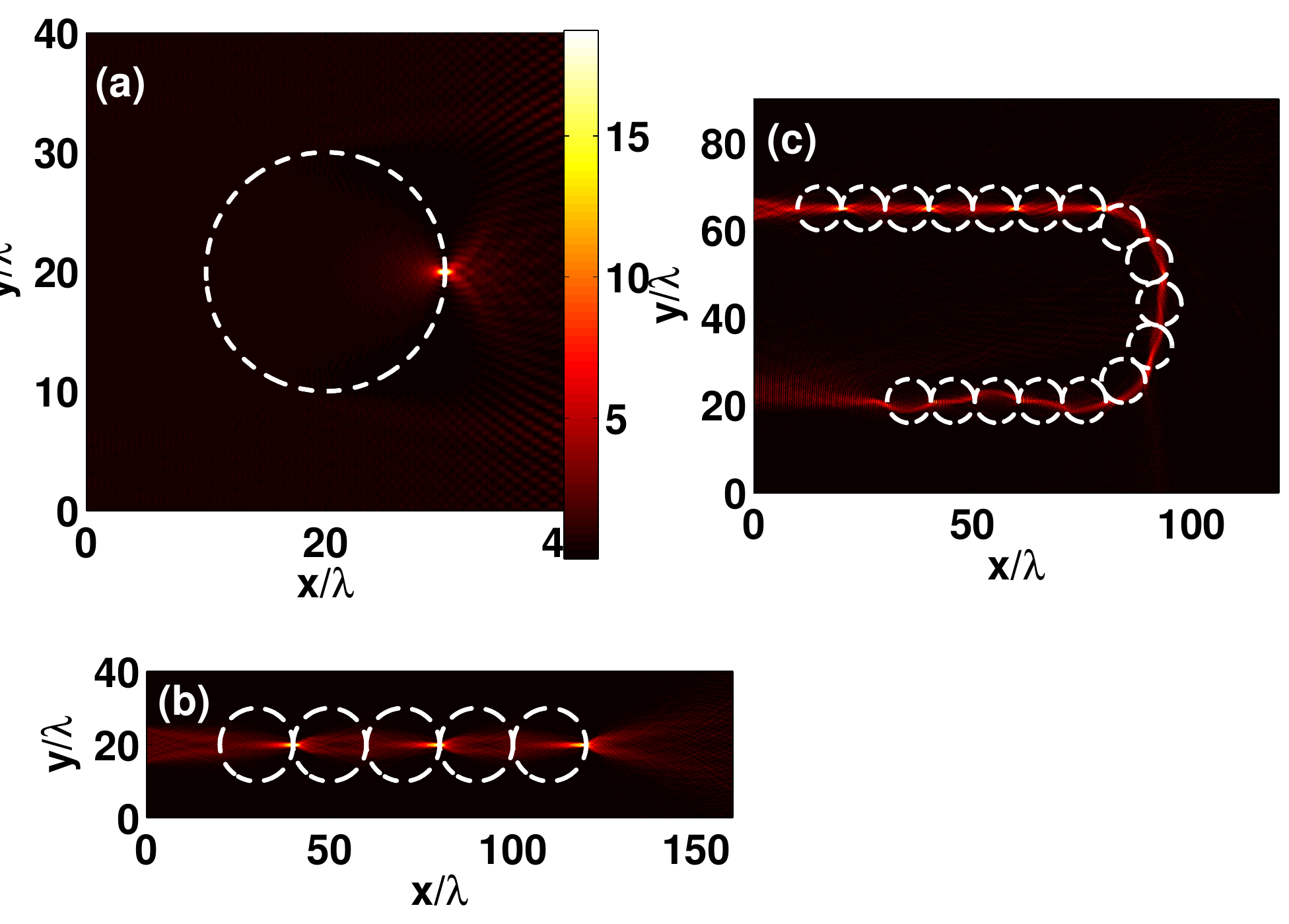}
 \caption{The white  dashed lines denote the arrangement of Luneburg lenses.  We present the intensity $I$ of an electromagnetic wave  propagating through (a) a single Luneburg lens, (b) a linearly arranged Luneburg lens waveguide formed by five Luneburg lenses, (c)  a $180^o$ reversed bend waveguide formed by seventeen Luneburg lenses. The simulations have been performed using the Finite Difference in Time Domain method.}
\label{fig:fig3}
\end{center}
\end{figure}

 Fig. \ref{fig:fig3} presents the steady state intensity of the electric field calculated by means of the FDTD \rcgindex{\myidxeffect{F}!Finite Difference in Time Domain} simulations, verifying the  analytical results that are shown in Figs. \ref{fig:fig1} and \ref{fig:fig2}.

In conclusion, in Section \ref{marios:sec:mathTools}, four methods have been discussed in order to investigate the light propagation through media with space-dependent  refractive indices $n(r)$ (equivalently, via media with permittivity $\varepsilon(r)$); namely, the Finite Difference in Time Domain for solving the time dependent Maxwell equations and  three geometrical optics (or ray tracing) methods  (the first two  are based on Fermat principle whereas the third method is based on Helmholtz wave equation). 

All the methods have been applied to investigate the electromagnetic wave propagation through  certain configurations of Luneburg lenses (a metamaterial gradient refractive index spherical lens with focus properties and index of refraction  given by the equation (\ref{marios:eq:luneburgIndex})). We have shown that all these methods are in agreement. In addition,  we  have discussed the formation of  waveguides which are formed by arrangement of multiple Luneburg lenses, in geometrically linear or bent configurations, resulting to enhanced control of light propagation.

\section{Branching flow in weakly disordered media}
\label{marios:sec:branching}

When waves propagate through random media many interesting phenomena occur, with the most well known being that of Anderson localization.  Among those, there are coherent phenomena related to branching\rcgindex{\myidxeffect{B}!Branching} of 
waves, the onset of caustic areas\rcgindex{\myidxeffect{C}!Caustics} as well as rogue wave formation\rcgindex{\myidxeffect{R}!Rogue waves}. Of particular interest are phenomena related to  electron flow in a  two dimensional electron gas  \cite{jakob2010,topinka2001}, transport properties of semiconductors \cite{jakob2010,topinka2001},  ocean waves \cite{heller}, linear and nonlinear light propagation in random fibers \cite{foivos,moti2,moti2013}, sound wave propagation \cite{benon, wolfson_tappert, wolfson_tomsovic}, microwave devices \cite{jakob2,hohmann2010}, resonance in nonlinear optical cavities \cite{montina} and  light propagation through random refractive index media \cite{klyatskin1993,marios2014,xuan,soli}. Many of these cases can be analyzed mathematically using a unified framework that provides results valid in quite different circumstances.
 
In this section, we focus on  branching effects that occur in two dimensional conservative particle flows through a weak random potential\rcgindex{\myidxeffect{R}!Random potential}. Even if the potential is very weak, the flow can be strongly influenced by the
disorder resulting in the onset of caustics \rcgindex{\myidxeffect{C}!Caustics} branches \cite{kaplan,jakob2010}. We first present the theoretical framework that has  been developed for the quantification of  branching\rcgindex{\myidxeffect{B}!Branching} effects  in the two-dimensional electron flow. We then  show that caustics emerge in an analogous way in the  propagation of light through a disordered network of lenses and we outline the similarities between light propagation and  electron flow.

\subsection{Statistics of caustics}
\label{marios:subsec:causticStatistics}

We present the theoretical framework for caustics  based on the Lagrangian manifold  approach  in order to obtain analytical results for the caustics statistics. The Lagrangian manifold approach offers the opportunity to adequately understand the phase space geometry of a caustic. The analytical results are general and  hold for a variety of problems,  since the initial point of this analysis is the Hamiltonian formalism \cite{jakobThesis,jakob2010}. An appropriate way to study the branched flow is to analyse the statistics of caustics, since each caustic is followed by  branched flow.   

We start with a Hamiltonian of the form
\begin{equation}
\label{marios:eq3:ordinaryHamiltonian}
H=\frac{\vec p^2}{2m}+V(t,\vec x)
\end{equation}
where  $\vec x$, $\vec p$  is the position and momentum vectors respectively of a particle while $t$ is time. 
The corresponding Hamilton-Jakobi  equation is a first order non-linear partial differential equation given by 
\begin{equation}
\label{marios:eq3:hje0}
\frac{\partial}{\partial t} S(t,\vec x) + H=0
\end{equation}
 where $S(t,\vec x)$ is the classical action which is associated with the conjugate momentum vector as
\begin{equation}
\label{marios:eq3:action}
\vec p(\vec x) = \frac{\partial S(\vec x)}{\partial \vec x}
\end{equation}
Substituting equation (\ref{marios:eq3:action}) and equation (\ref{marios:eq3:ordinaryHamiltonian}) in equation (\ref{marios:eq3:hje0})  and assuming particles with unit mass $m=1$ we obtain 
\begin{equation}
\label{marios:eq3:hje}
\frac{\partial}{\partial t} S(t, \vec x) + \frac{1}{2} \left( \frac{\partial S}{\partial \vec x} \right) ^2+V(t,\vec x) =0
\end{equation}
For ``weak" potentials  we can use the quasi two dimensional \rcgindex{\myidxeffect{Q}!Quasi 2D approximation} or the paraxial approximation with only one spatial coordinate, viz. $\vec x(t) =y(t)$ and $\vec p (t) = p(t)$ and with time $t$ playing the role of the propagation axis. The mathematical problem is that of  an (1+1) dimensional Hamilton-Jakobi equation with a time dependent potential $V\left(t,y(t)\right)$.

The  curvature $u$ of the action $S$ is defined as the partial derivative of conjugate momentum $p$ with respect to position $y$, namely
\begin{equation}
\label{marios:eq3:curvature}
u\equiv\frac{\partial p}{\partial y}=\frac{\partial^2 S}{\partial y^2}
\end{equation}
In order to obtain a differential equation for the curvature $u$, we  differentiate twice the  equation (\ref{marios:eq3:hje}) with respect to position $y$ and, by using the definitions of equations (\ref{marios:eq3:action})(\ref{marios:eq3:curvature}) \cite{klyatskin1993, jakobThesis}, we obtain
\begin{eqnarray}
\frac{\partial}{\partial t} u + \frac{\partial S}{\partial y}\frac{\partial}{\partial y}u +u^2+\frac{\partial^2}{\partial y^2} V(t,y)=0 \nonumber \\
\label{marios:eq3:calculations}
\left[ \frac{\partial}{\partial t} + p\frac{\partial}{\partial y}\right] u + u^2 +\frac{\partial^2}{\partial y^2} V(t,y)=0
\end{eqnarray}
The operator in the bracket of equation (\ref{marios:eq3:calculations}) is called convective or material derivative \cite{jakobThesis, ockendonBook}, turning the differential equation from a partial  to an ordinary one and, thus,  the Eulerian into a Lagrangian framework.

Equation (\ref{marios:eq3:calculations}) can take the form
\begin{equation}
\label{marios:eq3:mainODE}
\frac{d}{dt} u + u^2 +\frac{\partial^2}{\partial y^2} V(t,y)=0
\end{equation}
The next step is to introduce  random noise\rcgindex{\myidxeffect{R}!Random noise}. Since we are interested in  in wave propagation through weak random  potential\rcgindex{\myidxeffect{R}!Random potential}, we assume that the potential is simply
white noise  $\Gamma(t)$ with  correlation  function $c(t-t')=\left\langle \Gamma(t)\Gamma(t')\right\rangle = 2\delta(t-t')$, i.e. the
noise is delta-correlated.  Due to the paraxial approximation used,  the noise term needs to act only in the   propagation direction $t$ \cite{jakob2,klyatskin1993,jakobThesis,jakob2010}.  The correlation function $c(t,y)$  of the stochastic term $\partial_{yy} V(t,y)$ of equation (\ref{marios:eq3:mainODE}) is
\begin{equation}
c(t-t',y-y')=\left\langle \partial_{yy}V(t,y)~ \partial_{y'y'} V(t',y') \right\rangle = \partial_{yy} \partial_{y'y'} ~c(t-t' , y-y') \nonumber
\end{equation}
\begin{equation}
\label{marios:eq3:cor0}
c(t-t',y-y')=2\delta(t-t')  \partial_{yy} \partial_{y'y'} c(y-y') 
\end{equation}
Although we assume that the random noise\rcgindex{\myidxeffect{R}!Random noise} $\Gamma(t)$ acts only in the propagation direction $t$, we would like  to also retain the characteristics of the random potential\rcgindex{\myidxeffect{R}!Random potential} in the transverse axis $y$. This can be achieved by keeping constant the integral over derivatives of the correlation function $c(y-y')$ in the following way: \cite{klyatskin1993,jakob2010,jakobThesis}
\begin{equation}
\label{marios:eq3:diffusionCoef}
\sigma^2=\frac{1}{2}\int_{-\infty}^\infty \left. \frac{\partial^4}{\partial y^4} c(t,y) \right| _{y=0} dt
\end{equation}
where $\sigma$ is the standard deviation of the potential. The constant coefficient $D$  will be identified  as the diffusion coefficient, related to the standard deviation $\sigma$ as
\begin{equation}
\label{marios:eq3:SigmaD}
D= 2 \sigma^2
\end{equation}
Thus, the ordinary differential equation  (\ref{marios:eq3:mainODE}) becomes an ordinary stochastic differential equation  viz.
\begin{equation}
\label{marios:eq3:mainOSDE}
\frac{du(t)}{dt}=-u^2(t) -\sigma~\Gamma(t)
\end{equation}
In the following, we will use the Fokker-Planck equation\rcgindex{\myidxeffect{F}!Fokker Plank Equation}, which is a partial differential equation describing the time evolution of the probability density function.  The latter is derived from an ordinary stochastic differential equation \cite{StochasticSpringer,jakobThesis,FPEspringer} and in the one dimension takes the form 
\begin{equation}
\label{marios:eq3:OSDEform}
\dot y(t)= f(y)+g(y)\Gamma(t)
\end{equation} 
where $f$ and $g$ are arbitrary functions of $y$ and $\Gamma$ is a Gaussian delta-correlated white noise. The corresponding Fokker-Planck equation for the density function $P(y,t)$ has the form
\begin{equation}
\label{marios:eq3:FPEform}
\frac{\partial}{\partial t}P(y,t)= \left[ -\frac{\partial}{\partial y} D^{(1)}(y,t)+\frac{\partial^2}{\partial y^2} D^{(2)}(y,t) \right] P(y,t)
\end{equation}
with drift and diffusion coefficients $D^{(1)}$ and $D^{(2)}$ respectively,  calculated via equation (\ref{marios:eq3:OSDEform}) according to the relations
\begin{equation}
\label{marios:eq3:driftForm}
D^{(1)}(y,t) = f(y)+g(y)\frac{\partial}{\partial y}g(y)
\end{equation}
\begin{equation}
\label{marios:eq3:diffusionForm}
D^{(2)}(y,t) = g^2(y)
\end{equation}
In addition to  the Fokker-Planck equation, one may also use the  backward Fokker Planck Equation \rcgindex{\myidxeffect{F}!Fokker Plank Equation} (equation (\ref{marios:eq3:BFPEform}), presented below), in which the space independent variable is a function of the initial position $y$. The main difference between the forward and  backward Fokker-Planck equation is  that in the former the initial value for the probability density is given, i.e. $P(y_0,t_0)$ and the equation describes  the  time evolution of this density $P(y,t)$ for time $t>t_0$. On the other hand, in the backward Fokker-Planck equation,  the final condition  $P(y_f,t_f)$ is given, where $y_f$, $t_f$ are the final values of variables $y$ and $t$, while the  initial conditions are unspecified. The backward Fokker-Planck equation is very useful for the solution of problems where  the final state of process is known but we are not interested in or do not know the initial conditions. In order to avoid confusion, we use  $P$ for the probability density in the forward  and $p_f$ for the backward Fokker-Planck equations respectively; for the latter we have: 
\begin{equation}
\label{marios:eq3:BFPEform}
\frac{\partial}{\partial t_0} p_f(y,t) = \left[ -D^{(1)}(y_0,t_0)\frac{\partial}{\partial y_0}+D^{(2)}(y_0,t_0)\frac{\partial^2}{\partial y_0^2} \right]p_f(y,t)
\end{equation}
We derive the drift and the diffusion coefficient, based on equations (\ref{marios:eq3:OSDEform}), (\ref{marios:eq3:driftForm}) and (\ref{marios:eq3:diffusionForm}) for  equation (\ref{marios:eq3:mainOSDE})
\begin{equation}
\label{marios:eq3:drift}
D^{(1)}=-u^2
\end{equation}
\begin{equation}
\label{marios:eq3:diffusion}
D^{(2)}=\sigma^2 = \frac{D}{2} 
\end{equation}
The Fokker-Planck equation of our problem is given by equations (\ref{marios:eq3:FPEform}), (\ref{marios:eq3:drift}) and (\ref{marios:eq3:diffusion}), viz.
\begin{equation}
\label{marios:eq3:fpe}
\frac{\partial}{\partial t}P(u,t)= \left[ \frac{\partial}{\partial u}u^2 +\frac{\partial^2}{\partial u^2} \frac{D}{2} \right] P(u,t)
\end{equation}
In order to find the time necessary for the onset of a caustic\rcgindex{\myidxeffect{C}!Caustics} for the first time, viz. when the solution of the Fokker-Planck equation becomes infinity for the first time ($u(t_c)\rightarrow \infty$, where $t_c$  is the mean time of this process), we  ask the inverse question, i.e.  what is the probability that no singularity appears until  time $t$, (meaning that when a singularity appears, the process is terminated). This analysis can be performed by means of the backward Fokker-Planck equation\rcgindex{\myidxeffect{F}!Fokker Plank Equation} \cite{klyatskin1993,jakobThesis}. Using the form of equation (\ref{marios:eq3:BFPEform}) with coefficients given by equations (\ref{marios:eq3:drift}) and (\ref{marios:eq3:diffusion}), we have
\begin{equation}
\label{marios:eq3:bfpe}
\frac{\partial}{\partial t} p_f(u,t) = \left[ -u^2_0\frac{\partial}{\partial u_0}+\frac{D}{2} \frac{\partial^2}{\partial u^2_0} \right]p_f(u,t)
\end{equation}
where $u_0$ the initial curvature.

We proceed with the calculation of the mean time $\langle t_c(u_0)\rangle$ necessary for the initial curvature $u_0$ to diverge
and thus  produce a caustic. According to basic probability theory, the mean time $\langle t_c(u_0)\rangle$ is given by the probability density $p_f$ via the relation
\begin{equation}
\label{marios:eq3:meanDefinition}
\langle t_c(u_0)\rangle = \int_0^\infty t p_f ~dt 
\end{equation}
In order to calculate  $\langle t_c(u_0)\rangle$,  we multiply   the backward Fokker-Planck equation  (\ref{marios:eq3:bfpe}) by $t$ and afterwards we integrate over time $t$. The left hand side can be evaluated by means of the integration by parts method, resulting in
\begin{equation}
\label{marios:eq3:leftPart}
\int_0^\infty t \frac{\partial}{\partial t}~ p_f~ dt= \left. tp_f \right|_0^\infty - \int_0^\infty  p_f~ dt=0-1=-1
\end{equation}
We have assumed that the probability density $p_f$ is normalized to unity, that is	 $\int_0^\infty p_f =1$, and furthermore, it vanishes as time approaches infinity resulting in $p_f(t\rightarrow \infty)=0$. The left hand side does not include derivatives with respect to $t$ and, therefore, the integration is trivial; the equation thus becomes
\begin{equation}
\label{marios:eq3:tcDiff}
-1=-u^2_0\frac{d}{du_0} \langle t_c(u_0)\rangle+\frac{D}{2}\frac{d^2}{du_0^2} \langle t_c(u_0)\rangle
\end{equation}
where we have  used the definition of equation (\ref{marios:eq3:meanDefinition}) and transformed the partial derivatives (with respect to $u_0$) to full derivatives, since the time derivatives vanish. Equation (\ref{marios:eq3:tcDiff}) is a second order inhomogeneous differential equation of the form
\begin{equation}
y''(x)+f(x)~y'(x)=g(x) \nonumber
\end{equation}
with exact solution given by \cite{polyanin}
\begin{equation}
\label{marios:eq3:integralSolution}
y(x)=C_1+\int e^{-F}\left(C_2+\int e^{F}gdx\right)dx\quad \text{where}\quad F=\int fdx \
\end{equation}
Using equation (\ref{marios:eq3:integralSolution}) along with the boundary conditions 
\begin{equation}
\lim_{u_0 \to -\infty}\langle t_c(u_0)\rangle=0 \quad \text{and} \quad \lim_{u_0 \to \infty}\langle t_c(u_0)\rangle=\text{finite} 
\end{equation}
we obtain the final solution for the mean time  $\langle t_c(u_0)\rangle$ in terms of a double integral form, that is
\begin{equation}
\label{marios:eq3:tcFinalIntegral}
\langle t_c(u_0)\rangle= \frac{2}{D}\int_{-\infty}^{u_0} e^{2\xi^3/3D}\int_\xi^\infty e^{-2\eta^3/3D}d\eta d\xi
\end{equation}
The integral in equation (\ref{marios:eq3:tcFinalIntegral}) can be evaluated numerically for a plane wave or  point source condition, viz. $u_0=0$ and $u_0=\infty$ respectively, returning a numerical value for the characteristic mean time (or, equivalently, the distance in the quasi two dimensional  approximation\rcgindex{\myidxeffect{Q}!Quasi 2D approximation}) from a plane or from a point source, where the first caustic\rcgindex{\myidxeffect{C}!Caustics} appears \cite{klyatskin1993,jakobThesis}, namely
\begin{equation}
\langle t_c(0)\rangle = 4.18 D^{-1/3} \quad \text{and} \quad \langle t_c(\infty)\rangle = 6.27 D^{-1/3}\nonumber
\end{equation}
Employing equation (\ref{marios:eq3:SigmaD}) we can rewrite the results in terms of the standard deviation $\sigma$ as
\begin{equation}
\label{marios:eq3:tcResults}
\langle t_c(0)\rangle = 3.32 \sigma^{-2/3} \quad \text{and} \quad \langle t_c(\infty)\rangle = 4.98 \sigma^{-2/3}
\end{equation}

Note that we can derive the same results as above if we start from the parabolic equation  \begin{equation}
\label{marios:eq3:parabolicEq}
2ik\frac{\partial}{\partial t}\psi+\nabla^2 \psi+k^2 \epsilon(t,\vec r)\psi=0
\end{equation}
We can then use as a starting point the  Schr{\"o}dinger-like equation (\ref{marios:eq3:parabolicEq}), valid in the paraxial approximation,  instead of starting with the Hamiltonian of equation  (\ref{marios:eq3:ordinaryHamiltonian}) \cite{klineBook,klyatskin1993,stauroudisBook}. In this case, the time $t$ is also  the propagation axis (as in the paraxial approximation), $\psi=\psi(t,\vec r)$ is any component of electric or magnetic field, $k$ is the wavevector and $\epsilon$ is the dielectric coefficient. In this case, the  classical action $S$, which is defined by equation (\ref{marios:eq3:action}), is the phase front of the electromagnetic wave, and the curvature $u$ denotes the curvature of the phase front; $\epsilon$ is the random potential\rcgindex{\myidxeffect{R}!Random potential} (random fluctuated permittivity) \cite{klyatskin1993}.

These results prove that  the onset of caustics \rcgindex{\myidxeffect{C}!Caustics} is a general phenomenon taking place in conservative particle flows as well as in wave propagation through a weak delta-correlated random potential. We have shown that the characteristic mean distance from the source, where the first caustic occurs, is universal for all such systems and it is given in terms of standard deviation of the random potential according to the relations (\ref{marios:eq3:tcResults}).

 In the following subsections we present numerical results that  have been obtained through simulations of particle flows and wave propagation systems; the numerical results are in agreement with the analytical results presented in this subsection (\ref{marios:subsec:causticStatistics}).

\subsection{Branching flows in physical systems}
\label{marios:subsec:causticsInPhysical}

Numerical simulations as well as experiments have revealed that branching\rcgindex{\myidxeffect{B}!Branching} flows can arise in a variety of physical systems.  Topinka \textit{et al.} \cite{topinka2001} have shown experimentally that branching flow takes place in electron currents in a two dimensional electron gas.  Kaplan \cite{kaplan} and  Metzger  \cite{jakobThesis} have studied both analytically and numerically the branching flow in electron propagation and  have found that the scaling law governing the scaling behavior of the first caustic position is the one described in subsection \ref{marios:subsec:causticStatistics} (equation (\ref{marios:eq3:tcResults})). In addition,  Metzger \textit{et al.}, in \cite{jakob2010}, have found an analytical expression for the number of branches that occur in
various distances from the source. Moreover  Barkhofen \textit{et al.} \cite{jakob2} have found experimentally a branching effect in microwave flow through  disordered media fabricated though randomly distributed scatterers;  additionally, they have shown that the statics of the position of the first caustic satisfies the scaling rule described by equation (\ref{marios:eq3:tcResults}). Another microwave study that has been performed by  Hohmann \textit{et al.} \cite{hohmann2010}, found by means of the ray dynamics method and by wave propagation simulations, that branching flow can emerge in two dimensional microwave propagation through  media  comprising of random metallic scatterers.   Ni \textit{et al.} \cite{xuan} have studied the electromagnetic wave propagation in an optical system made of random scatterers with continuous refractive index, and have proposed that branched waves can emerge as a general phenomenon in the regime between the weak scattering limit and Anderson localization. In addition, they have found that high intensities (i.e. caustics) are distributed following an algebraic law. A numerical investigation on sound waves has also been performed by  Blanc-Benon \textit{et al.}  \cite{benon}  showing that branching flow can arise from high frequency sound wave propagation through a turbulent field; experiments performed by  Wolfson in \cite{wolfson_tomsovic} confirm the numerical findings.

\begin{figure}[h] \begin{center}
\includegraphics[scale=.45]{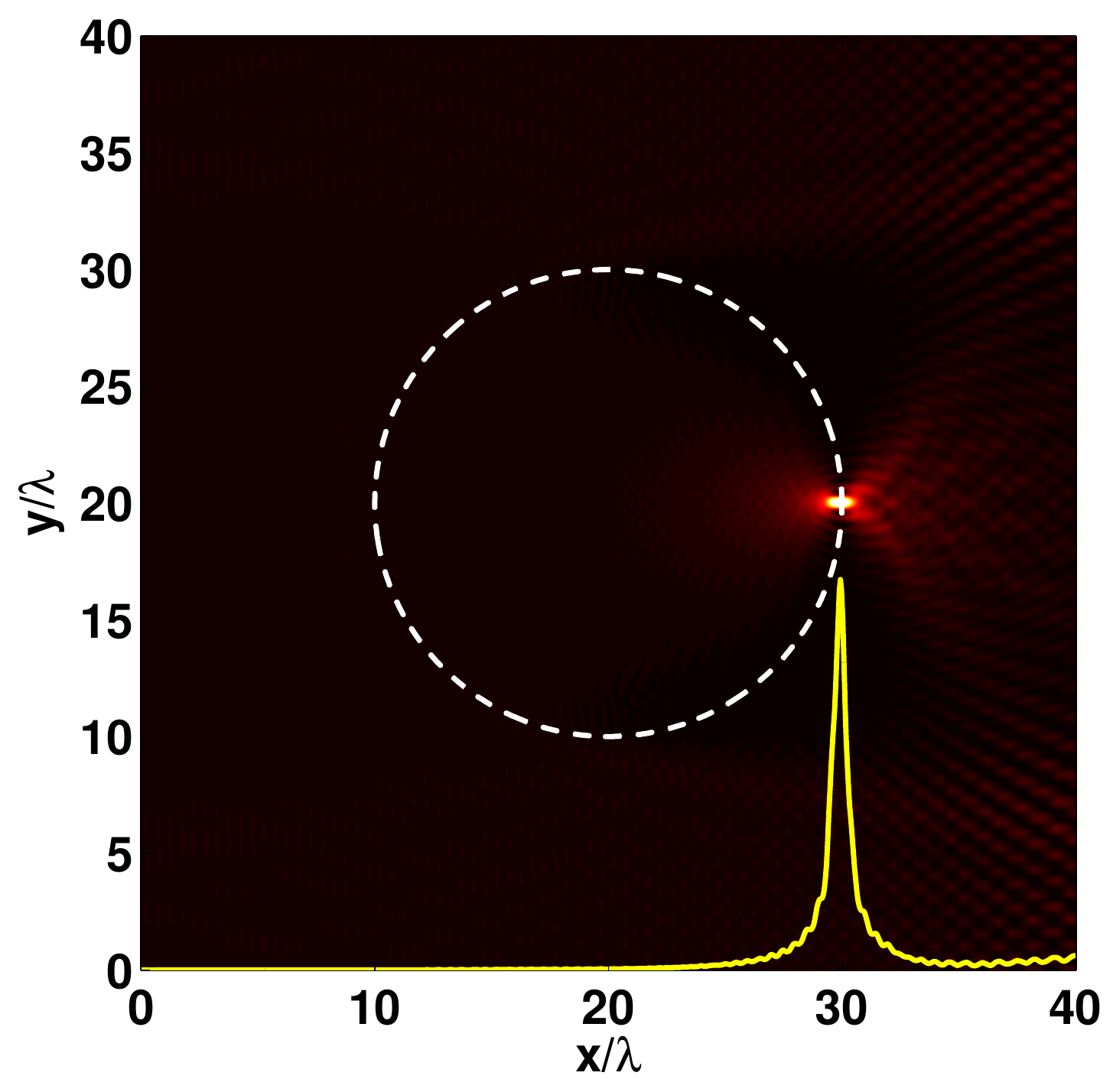}
 \caption{(Color online) Finite Difference in Time Domain simulation for electromagnetic wave propagation through a single Luneburg lens is illustrated; Luneburg lens is shown by the dashed white line, the lighter color denotes high intensity and the darker one is for lower intensity. The yellow solid line is the scintillation index, $\sigma_I^2$, given by equation (\ref{marios:sec3:scintillation}). As can been seen,  $\sigma_I^2$ takes its maximum value in the focus point.}
\label{fig:fig4b}
\end{center}
\end{figure}

The theoretical methods presented in subsection  \ref{marios:subsec:causticStatistics} will be now used in the specific  cases of  Luneburg lenses  (either in a single lens or  distributions of them) and we will investigate the location for  the occurrence of  the first caustic. The theoretical method will subsequently be compared to numerical results demonstrating
that  the latter  agree very well with the scaling law of equation (\ref{marios:eq3:tcResults}). As it has been already mentioned previously, while  the   wave intensity is very large in the caustic\rcgindex{\myidxeffect{C}!Caustics} regime and  substantial  deviations of the wave intensity are expected to appear.  In this regime, a maximum is  expected for the  standard deviation of the wave intensity $I$. 
A simple quantitative measure  to investigate the caustic formation  is  the scintillation index\rcgindex{\myidxeffect{S}!Scintillation index} \cite{andrews,jakob2}, which may be  studied as a function of the propagation distance $x$. When the average wave intensity $I$ is calculated using  many realizations of the random potential, the maximum of the scintillation index\rcgindex{\myidxeffect{S}!Scintillation index} depicted through a peak in the curve of $\sigma_I$) denotes the onset of a caustic. We note that  peaks for different values of the standard deviation $\sigma$ of the random potential\rcgindex{\myidxeffect{R}!Random potential} are scaled as equation (\ref{marios:eq3:tcResults}) predicts.  The scintillation index  $\sigma_I$ is defined as follows:
\begin{equation}
\label{marios:sec3:scintillation}
\sigma^2_I=\frac{\left\langle I(x)^2\right\rangle}{\left\langle I(x)\right\rangle^2}-1
\end{equation}
An alternative measure for the scintillation\rcgindex{\myidxeffect{S}!Scintillation index} index is obtained if we  average the intensity over the transverse direction $y$, viz.  
\begin{equation}
\label{marios:sec3:scintillationY}
s^2_I=\frac{\left\langle I(x)^2\right\rangle_y}{\left\langle I(x)\right\rangle_y^2}-1
\end{equation}
This definition is a more appropriate  when the average is taken over a few realizations random realizations of the medium.
In Fig. \ref{fig:fig4b},   we show that  the peak of scintillation index\rcgindex{\myidxeffect{S}!Scintillation index} $\sigma_I^2$ curve coincides with the location  where the electromagnetic wave is focused by the Luneburg lens, demonstrating that $\sigma_I$ is an efficient way to investigate caustics.

 We proceed with numerical simulations of electromagnetic plane wave propagation through a random transparent medium consisting of randomly located Luneburg lenses, each with refractive index profile given by equation (\ref{marios:eq:luneburgIndex}). The simulations utilize the Finite Difference in Time Domain \rcgindex{\myidxeffect{F}!Finite Difference in Time Domain} method, as it is described in subsection \ref{marios:subsec:fdtd}.

In order to investigate the branching\rcgindex{\myidxeffect{B}!Branching} flow for several values of the standard  deviation $\sigma$ of the random Luneburg lens potential\rcgindex{\myidxeffect{R}!Random potential}, we generalize the original Luneburg index  by introducing a strength parameter $\alpha$ in the Luneburg lens refraction index function (\ref{marios:eq:luneburgIndex}); this control parameter $\alpha$ is proportional to the standard deviation, i.e. $\sigma \sim \alpha$. The generalized Luneburg lens refractive index function is then given by the equation
\begin{equation}
\label{marios:eq3:LLgeneralizedIndex}
n(r)= \sqrt{\alpha\left(n_L^2-1 \right)+1}
\end{equation}
where $n_L$ denotes the original Luneburg lens\rcgindex{\myidxeffect{L}!Luneburg lens} refraction index given by equation (\ref{marios:eq:luneburgIndex}). For $\alpha=1$ we obtain the original Luneburg lens index, while for $\alpha=0$ we have a flat refractive index ($n=1$).

For the simulations, a monochromatic   electromagnetic plane wave source of wavelength $\lambda$ and with transverse magnetic polarization (TM) has been located at the beginning (at the left side) of a rectangular lattice. A random network consisting of 150 randomly located Luneburg lenses each with radius $R=10\lambda$ has been used;  $\lambda$ has been used as a normalized unit of length.  The size of the disordered rectangular lattice is $460\lambda\times 360\lambda$ with a  constant filling factor $f=0.28$. We use periodic boundary conditions at the up and down edges  and absorbing boundary condition at the end. 

\begin{figure}[h] \begin{center}
\includegraphics[scale=.32]{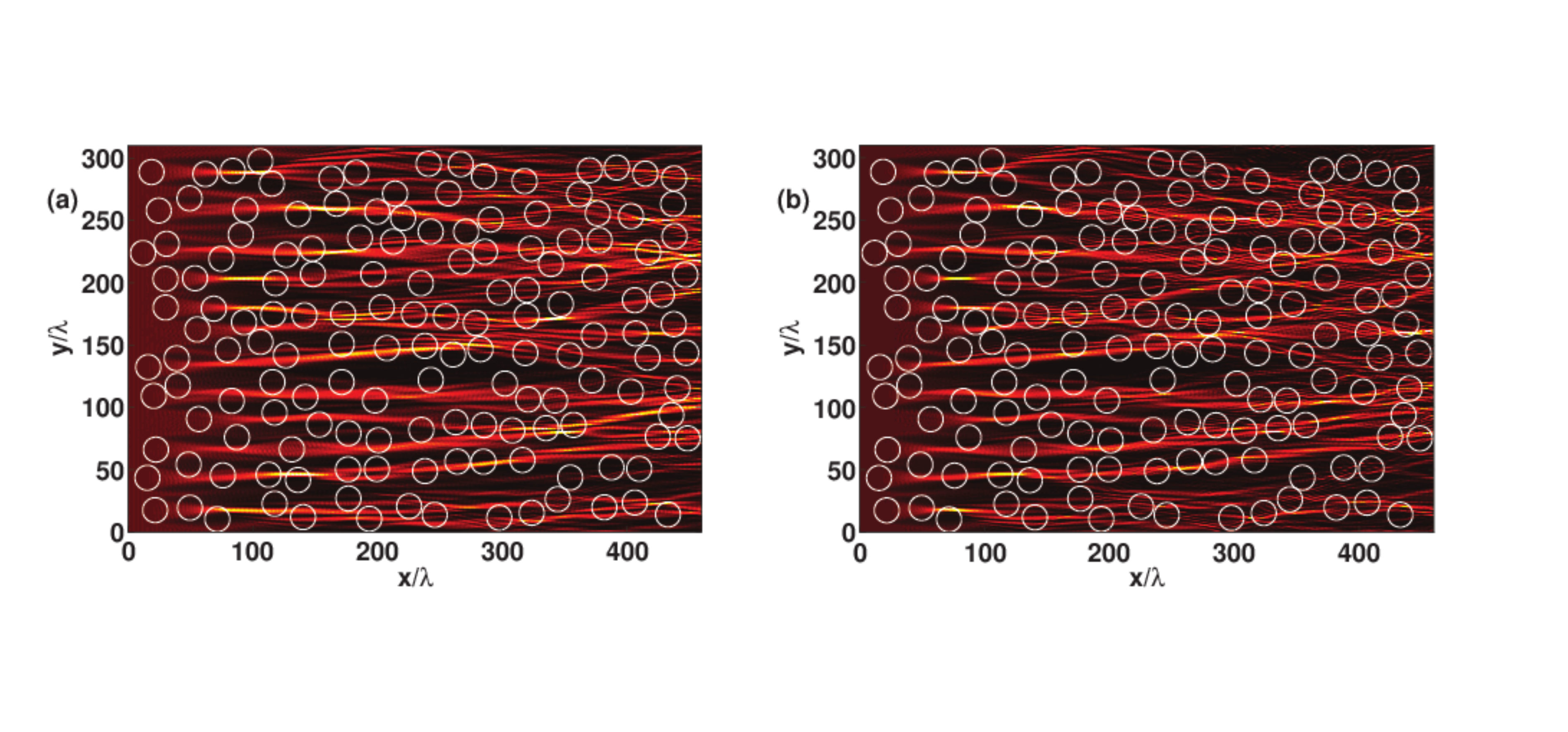}
 \caption{(Color online) White  lines denote the position of Luneburg lenses. Monochromatic electromagnetic plane waves propagate through a disordered transparent media consists of generalized Luneburg lenses with index of refraction given by equation (\ref{marios:eq3:LLgeneralizedIndex}). The intensity of electric field is denoted by lighter color for high intensity and by darker color for  lower intensity. In (a), the strength parameter is $\alpha=0.07$ while in (b) $\alpha=0.1$. In both images the branching flow is evident. }
\label{fig:fig5}
\end{center}
\end{figure}

The  intensity of the electric field  of the electromagnetic wave simulations through  random Luneburg lens networks for two different values of strength parameter $\alpha$ is represented in  Fig. \ref{fig:fig5}. The randomly located Luneburg lenses are illustrated by means of  white solid lines;  the lighter color denotes high intensity areas whereas the darker denotes lower values of intensity. Fig. \ref{fig:fig5}a indicates the propagation for $\alpha=0.07$, whereas  Fig. \ref{fig:fig5}b shows the  propagation for $\alpha=0.1$. 

\begin{figure}[h] \begin{center}
\includegraphics[scale=.3]{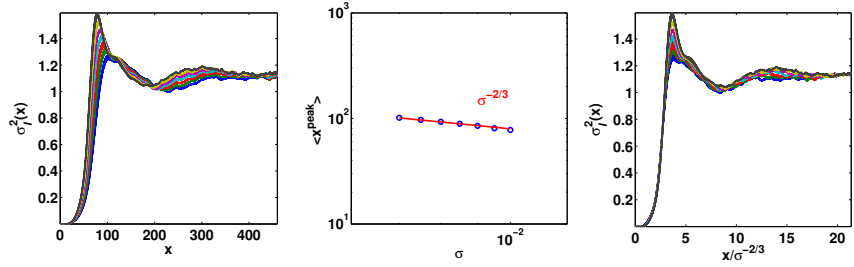}
 \caption{ Scaling of the branching length with respect to the standard deviation of the random potential $\sigma$. (a) Scintillation index $\sigma_I^2(x)$ given by equation (\ref{marios:sec3:scintillation}) as a function of the distance from the source, for different values of $\sigma$ (b) maximum position of the scintillation curves obtained from $\sigma_I^2$; the curve shows a scaling of $\sigma^{-2/3}$ (solid line). The scaling is confirmed in panel (c), where the curves from the first panel (a) are shown with a rescaled $x$ axis, viz. $x\rightarrow x/\sigma^{-2/3}$, in which all peaks occur at approximately the same distance. }
\label{fig:fig5b}
\end{center}
\end{figure}

 In Fig. \ref{fig:fig5b} we show the scintillation index\rcgindex{\myidxeffect{S}!Scintillation index} $\sigma_I^2$, as  it is given by equation (\ref{marios:sec3:scintillation}), for several values of strength parameter $\alpha$ (viz. several values of potential standard deviation $\sigma$). For each value of $\alpha$ we have computed the average of scintillation index for 300 simulations obtaining good accuracy. In Fig. \ref{fig:fig5b}a we plot  the  mean value of $\sigma_I^2$ as a function of the propagation coordinate $x$, whereas in  Fig. \ref{fig:fig5b}c the same curves are illustrated in a rescaled $x$ axis, i.e. $x\rightarrow x/\sigma^{-2/3}$. The position $x_{peak}$ where the maximum of scintillation index curves  is found for each standard deviation (or each $\alpha$), are plotted  in the Fig. \ref{fig:fig5b}b; the slope of the solid line  in Fig.\ref{fig:fig5}a shows how these points are distributed, i.e. which is the relation between the first caustic position and the standard deviation of the random potential,  manifesting that the theoretical findings of equation (\ref{marios:eq3:tcResults}) holds  in the caustic formation of propagation of light through a random medium.

In conclusion, in Section \ref{marios:sec:branching}, we have developed a theory based on Hamilton-Jakobi equation for the description of propagation through disordered media,  assuming that randomness acts as a white noise upon the flow. This assumption leads to a Fokker-Planck equation which is a well known partial differential equation derived by an ordinary stochastic differential equation, describing the time evolution of a probability density function. Afterwards, we have calculated the average distance (or the mean time in the paraxial approximation) from the source where the first caustic appears as a function of the standard deviation of random potential resulting to the scaling law given by the equation (\ref{marios:eq3:tcResults}). Furthermore, we proceeded with numerical simulations for monochromatic electromagnetic plane waves through a disordered lattice consisting of randomly located generalized Luneburg lenses with refractive index given by the equation (\ref{marios:eq3:LLgeneralizedIndex}); these simulations are taken place for several values of standard deviation (or Luneburg lens strength parameter $\alpha$). In addition we introduced the scintillation index $\sigma_I$, which is a useful quantity for determining a caustic. The numerical results prove the validity of the theory that is discussed, since they show that the relation between the first caustic position and the standard deviation of the random potential is in agreement with the theoretical findings.  It should be mentioned that it is a very interesting fact that the branched flow effect and the law that rules the position of the first caustic in a turbulence flow is the same for conservative particle flow as well as for the wave propagation in random media.

\section{Rogue wave formation through strong scattering random media}
\label{marios:sec:RWs}

Rogue waves\rcgindex{\myidxeffect{R}!Rogue waves}  or freak waves, have for long triggered the interest of scientists because of their intriguing properties. They are extreme coherent waves with very large magnitudes,  which appear suddenly from nowhere and disappear equally fast. Rogue waves were first documented  in relatively calm water in the open seas \cite{heller2,heller} but recent works have demonstrated  that rogue wave-type extreme events\rcgindex{\myidxeffect{E}!Extreme events} may appear in various  physical systems such as microwaves, nonlinear crystals, cold atoms and Bose-Einstein condensates, as well as in non-physical systems such as financial systems \cite{akhmediev2013,bacha2012,jakob2,hohmann2010, marios2014,montina,soli,wang2013,yan2010}.

Rogue wave pattern formation emerges in a complex environment but it is still unclear if their appearance is due to linear or nonlinear processes. Intuitively, one may link the onset of rogue wave pattern formation to a resonant interaction of two or more solitary waves that are present in the medium;  it has been tacitly assumed that extreme waves\rcgindex{\myidxeffect{E}!Extreme events} are due to nonlinearity \cite{bacha2012,maluckov2009,maluckov2013,montina,soli, yan2010}. However, large amplitude events may also appear in a purely linear regime \cite{jakob2,heller2,hohmann2010,marios2014,heller}; a typical example is the generation of caustic surfaces in the linear wave propagation as it was discussed in Section \ref{marios:sec:branching}.
 
\begin{figure}[h] \begin{center}
\includegraphics[scale=.28]{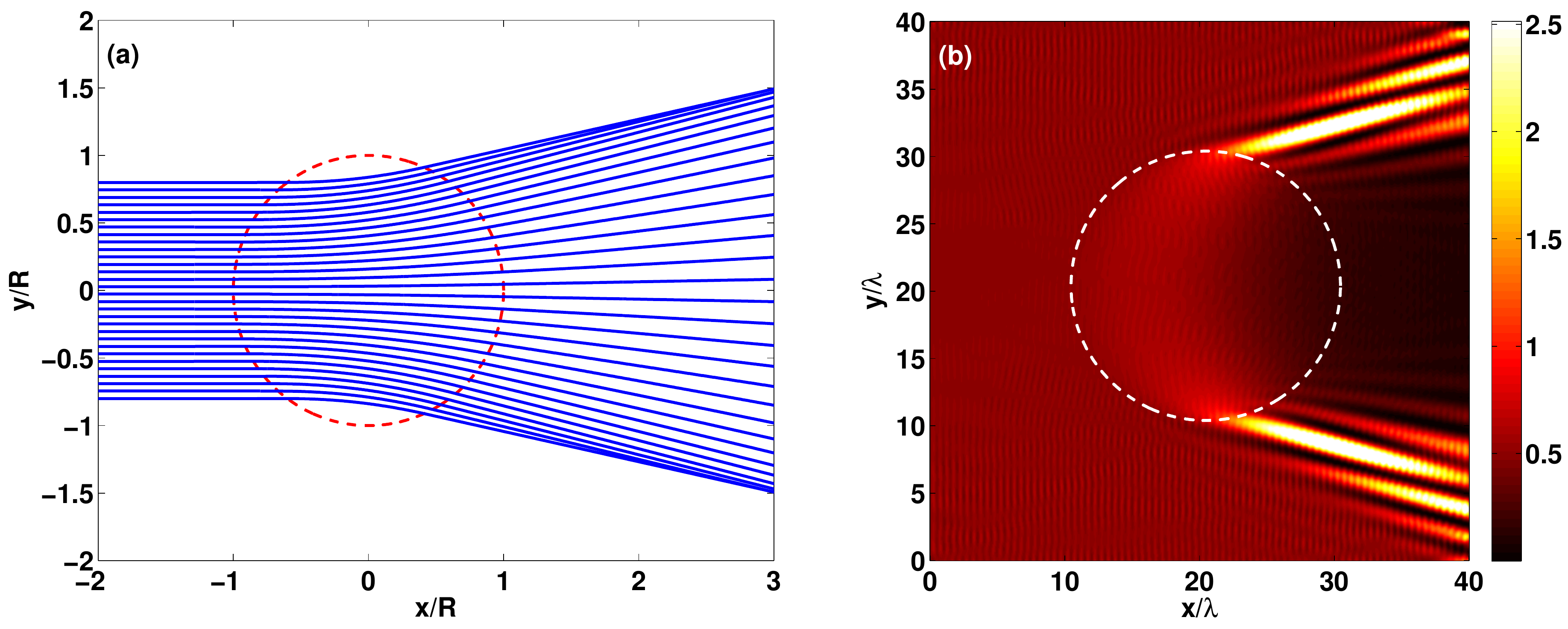}
 \caption{ The dashed lines  denotes   Luneburg hole  lenses, which are spherical lenses with index of refraction given by  equation (\ref{marios:eq4:LHindex}). In (a) we present an exact  solution of ray tracing propagation obtained by solving the ray equation (\ref{marios:eq:eqmotion2b}) or equation (\ref{marios:eq:rayTracing3}), with plane wave initial conditions while in (b) we present Finite Difference in Time Domain simulation results of monochromatic electromagnetic plane wave propagation through a single Luneburg hole lens, revealing the purely defocussing properties of Luneburg hole lens.}
\label{fig:fig6}
\end{center}
\end{figure}

In this Section we investigate optical wave propagation in  strongly scattering  optical media comprising Luneburg-type lenses randomly embedded in the bulk of transparent glasses. In particular, we use a type of lenses, namely Luneburg Holes \rcgindex{\myidxeffect{L}!Luneburg holes} (or anti-Luneburg lenses) instead of original Luneburg lenses with refractive index profile given by equation (\ref{marios:eq4:LHindex}) \cite{marios2014}. In contrast to a Luneburg lens,  the Luneburg hole has  purely defocussing properties as it is illustrated in Fig. \ref{fig:fig6}, where Fig.\ref{fig:fig6}a is ray tracing solution of ray equations  (\ref{marios:eq:eqmotion2b}) and (\ref{marios:eq:rayTracing3}) for the refractive index of equation (\ref{marios:eq4:LHindex}) whereas Fig.\ref{fig:fig6}b is a wave simulation performed by Finite Difference in Time Domain method. The difference of the index of refraction for Luneburg lenses  as well as holes, compared to the background index, is very large, viz. of the order of $40\%$ and thus a medium with a random distribution of Luneburg holes can be characterized as a strongly scattering random media. We are using this kind of lenses instead of original Luneburg lenses, because they are easier to be fabricated in the bulk of a dielectric, such as  silica glass \cite{marios2014}. 

By analysing the electromagnetic wave propagation  in the linear regime we observe the appearance of rogue type waves\rcgindex{\myidxeffect{R}!Rogue waves} that depend solely on the scattering properties of the medium. Numerical simulations have been performed using the Finite Difference in Time Domain\rcgindex{\myidxeffect{F}!Finite Difference in Time Domain} method, as it was discussed in subsection \ref{marios:subsec:fdtd}, showing that optical rogue waves are generated through  strong scattering in such a complex environment.
\begin{equation}
\label{marios:eq4:LHindex}
n(r)=\sqrt{1+\left(\frac{r}{R}\right)^2}
\end{equation}

\subsection{Rogue waves in optics}
\label{marios:subsec:RWsOptics}

As it has been already mentioned, rogue waves\rcgindex{\myidxeffect{R}!Rogue waves} are extreme coherent waves with very large magnitude; a more precise definition of rogue waves specifies that the height or intensity of a rogue wave has to be at least two times larger than significant wave height\rcgindex{\myidxeffect{S}!Significant wave height}  $H_s$, where the latter is defined as the mean wave height  of the highest one third wave height distribution \cite{heller2,marios2014,heller}.

Another way to study rogue waves is by means of the distribution of wave heights or intensities. According to the central limit theorem and the simple random wave prediction for the probability  distribution of wave intensities $I$, the intensities have to follow the Rayleigh law, obeying a distribution of  $P(I)=e^{-I}$, where $I=|E|^2$ (E is the electric field), normalized to one. However, when extreme events appear, the intensities distribution deviates from simple exponential and long tails appear, due to the presence of very high intensities \cite{heller2,hohmann2010,marios2014,heller}.

In the following, we present Finite Difference in Time Domain \rcgindex{\myidxeffect{F}!Finite Difference in Time Domain} numerical simulations for the electromagnetic wave propagation through media consisting of random located Luneburg holes. Each Luneburg hole lens, with refractive index given by equation (\ref{marios:eq4:LHindex}), has radius $R=3.5\lambda$, where $\lambda$ is the wavelength of the electromagnetic wave. The medium has dimensions $(175.0 \times 528.5)$ in ($\lambda^2$ units) and 400 Luneburg hole lenses are placed randomly in the dielectric (medium) with fixed filling factor $f=0.17$; absorbing boundary conditions have been applied.

\begin{figure}[h] \begin{center}
\includegraphics[scale=.35]{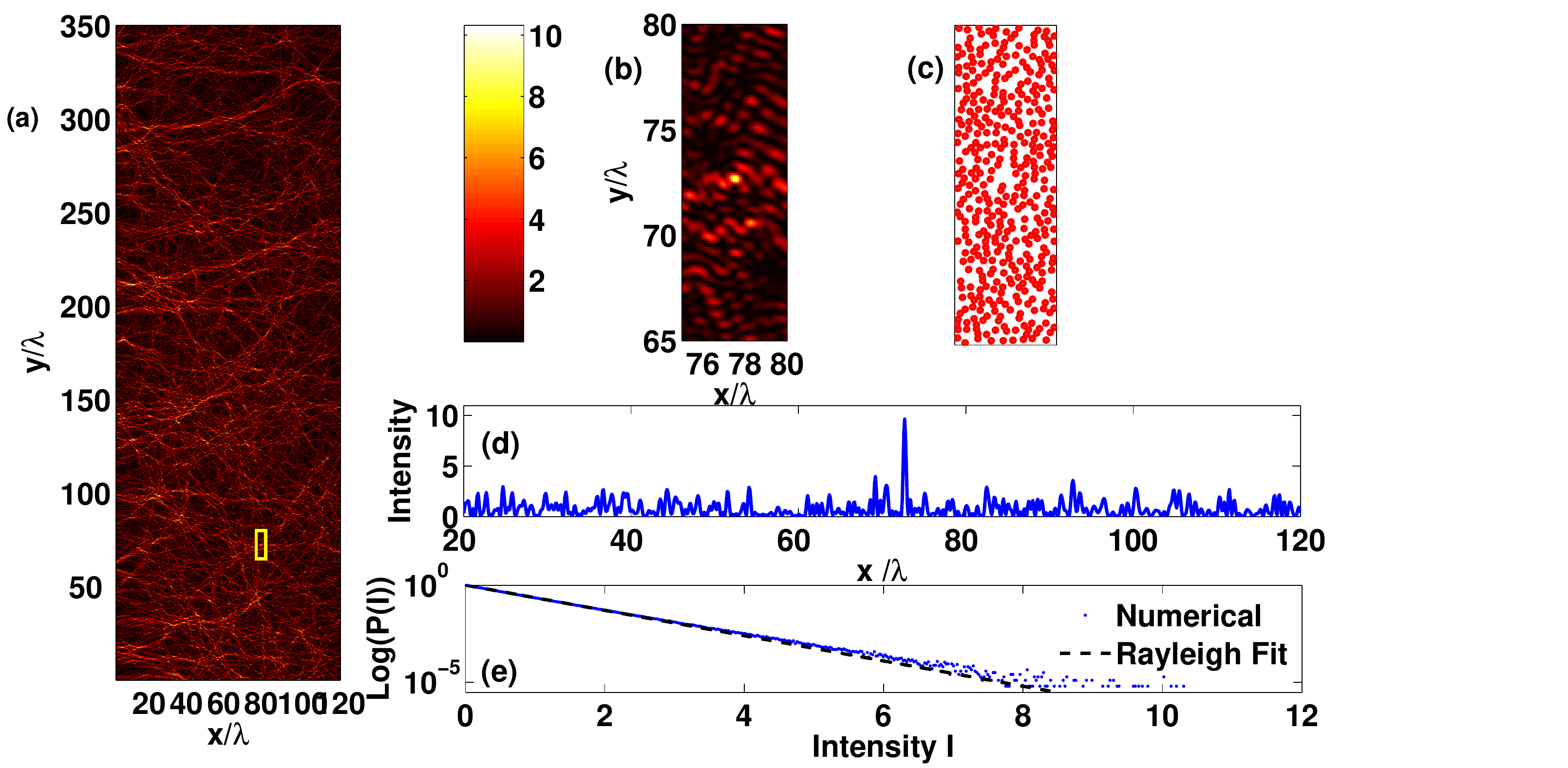}
 \caption{(Color online) (a) A monochromatic plane wave beam propagates, from left to right, through the Luneburg hole lenses disordered lattice. (b) A detail of the propagation (yellow box in (a)) showing an optical rogue wave-type. (c) A two dimensional random Luneburg hole network used in the simulations; each red circle represents a Luneburg hole lens. (d) Intensity profile in the rogue wave region as a function of $x$. (e) Intensities distribution (blue dots) (in semilog scaling) for the entire lattice shows deviation from the Rayleigh curve (black dashed line) resulting in an extreme event signature.}
\label{fig:fig7}
\end{center}
\end{figure}

In Fig. \ref{fig:fig7} we present the numerical results based on the Finite Difference in Time Domain method for the linear medium. In Fig. \ref{fig:fig7}a and in Fig. \ref{fig:fig7}b,  we present the propagation of a monochromatic electromagnetic plane wave with transverse magnetic (TM) polarization, through a random Luneburg hole network, where a plane wave source has been located on the beginning of lattice (left) and the wave propagates from the left to the right direction. We observe that the presence of scatterers with strong defocussing properties forces light to form propagation channels (Fig. \ref{fig:fig7}a) that can lead to  the generation of very large amplitude rogue type waves (Fig. \ref{fig:fig7}b). In Fig. \ref{fig:fig7}c, the random Luneburg hole network which is used for Finite Difference in Time Domain simulation of Figs \ref{fig:fig7}a and \ref{fig:fig7}b, is presented. Fig. \ref{fig:fig7}d shows  the intensity profile where a linear rogue type wave\rcgindex{\myidxeffect{R}!Rogue waves} occurs; as can been seen, the highest pick is larger than twice the significant wave height\rcgindex{\myidxeffect{S}!Significant wave height} resulting in a rogue wave. Fig. \ref{fig:fig7}e represents, in semilog axis, the distribution of electric intensities (blue dots) and the Rayleigh distribution (dashed black line). As can be seen,  the distribution of intensities deviates from the Rayleigh curve resulting in an extreme event signature.

Concluding, rogue  waves are extreme waves that appear in diverse systems; in Section \ref{marios:sec:RWs}, we have focused on studying complex media in which randomly placed elements introduce strong light scattering and
interference patterns. In the purely linear regime, the coalescence of these light
channels and the resulting complexity leads to the appearance of extreme, transient
waves. In addition to high intensity profile, there is a clear departure from the Rayleigh law in large intensities where rogue waves are produced, as a result we have a clear signature of extreme events.  The most important  result of this Section is that optical extreme events are generated in strong scattering linear media by the complexity of the medium that drives interference and wave coalescence.

\section{Conclusion}
\label{marios:sec:conclusion}

We have presented mathematical methods and tools in order to investigate electromagnetic wave propagation in optical media by means of ray tracing and the Finite Difference in Time Domain methods. We have shown that the ray tracing equations which have been developed in Section \ref{marios:sec:mathTools} (subsections \ref{marios:subsec:quasi2D}, \ref{marios:subsec:rayParametric} and \ref{marios:subsec:helmholtzRays}) can be used for the derivation of an exact ray equation for a given refractive index function, as well as for simulations solving  numerically the ray tracing equation of motion. We have verified these results by comparing them to the results obtained by the Finite Difference in Time Domain simulations.

After presenting the mathematical tools for wave propagation, we  investigated the limits of a disordered optical lattice which consist of gradient refractive index lenses with specific refractive index profiles. We  investigated the weak scattering limit in random media consisting of weak generalized type Luneburg lenses. We found that caustic formation and, therefore, branching flow can arise in such a system. We also investigated the strong scattering limit by simulating the propagation of monochromatic electromagnetic plane waves in  a random optical network which consists of Luneburg Hole lenses at random locations. We found that rogue wave formation can arise in such a system  even with the absence of nonlinearity. 

The appearance of branching flows and extreme waves of rogue type in 
disordered optical media shows that complexity inherent in the latter systems leads
to forms of self-organization:  branching introduces pathways for
light propagation in the medium; rogue waves result from extreme focusing and
coherence. These features show that the medium, especially in the strong scattering regime,
exhibits collective properties that emerge as a result of light scattering and propagation.
It is very interesting to continue this line of research in the direction of establishing
effective equations that describe the complex dynamics of light propagation in strongly
scattering, disordered media.  At the same time it is important to understand the
onset of rogue waves in these media since this will lead to possible control of the
generation and detection of these waves.  Research in complex optical media will 
undoubdely lead to numerous new and promising technology applications.

\section*{Acknowledgments}

We acknowledge  useful discussions with J.J. Metzger, R. Fleischmann and G. Neofotistos.
This work was supported in part by the European Union program FP7-REGPOT-2012-2013-1 under grant agreement 316165.
\\ \mbox{}\\

\newcommand{\noopsort}[1]{} \newcommand{\printfirst}[2]{#1}
\newcommand{\singleletter}[1]{#1} \newcommand{\switchargs}[2]{#2#1}

\end{document}